\def\HI{H{\small I}\ }
\def\HII{H{\small II}\ }
\def\HInospace{H{\small I}}
\def\HIInospace{H{\small II}}
\def\simgt{\lower.5ex\hbox{$\; \buildrel > \over \sim \;$}}
\def\simlt{\lower.5ex\hbox{$\; \buildrel < \over \sim \;$}}

\documentclass[useAMS,usenatbib]{mn2e}

\usepackage{graphicx}

\title[The rotation curves in the gas of edge-on galaxies]
  {Structure and kinematics of edge-on galaxy discs --
III. The rotation curves in the gas}
\author[M. Kregel \& P.~C. van der Kruit]
  {M. Kregel$^1$ and 
  P~.C. van~der~Kruit$^1$\thanks{E-mail: vdkruit@astro.rug.nl}\\
  $^1$Kapteyn Astronomical Institute, University of Groningen,
  P.O.Box 800, 9700AV Groningen, the Netherlands}
\begin{document}

\date{Accepted. Received.}

\pagerange{\pageref{firstpage}--\pageref{lastpage}} \pubyear{2004}

\label{firstpage}

\maketitle

\begin{abstract}
A technique is introduced for deriving the gaseous rotation curves
  of edge-on spiral galaxies. The entire major axis position-velocity
  (XV) diagram is modeled with a set of rings in a least-squares
  sense, allowing for the effects of beam-smearing and line-of-sight
  projection. The feasibility of the technique is demonstrated by
  applying it to good quality \HI XV diagrams of eight edge-on
  spirals. For seven additional spirals the XV diagrams are of
  insufficient quality, and the \HI rotational velocities derived
  earlier using the envelope-tracing method are retained. The \HI
  results are augmented with the optical emission line (\HIInospace)
  kinematics to arrive at estimates of the full rotation curves. A
  detailed comparison of the \HI and \HII kinematics shows that
the discs in our sample are sufficiently transparent at the heights 
above the plane where we have taken our optical spectra to derive
the stellar kinematics.
In several of these spirals the \HII is mainly confined
  to spiral arms and does not extend out to the edge of the \HI layer,
which may have  caused the \HII velocity profiles to be 
significantly narrower than  those of the \HInospace.

This paper has been accepted by MNRAS and is available in pdf-format
at the following URL:\\

http://www.astro.rug.nl/$\sim $vdkruit/jea3/homepage/paperIII.pdf

\end{abstract}


\end{document}